\documentclass[twocolumn]{aastex62}

\usepackage{amsmath}
\usepackage{graphicx}
\usepackage{natbib}
\usepackage{txfonts}
\usepackage[referable]{threeparttablex}
\usepackage{tensor}
\usepackage{mathtools}
\usepackage{mathrsfs}
\usepackage{multirow}
\usepackage{bm}
\usepackage{booktabs}

\usepackage{natbib}
\shorttitle{Void Bias from SUS }
\shortauthors{Chan, Li, Biagetti, and Hamaus}

\newcommand{\MpcOh}{ \,  \mathrm{Mpc} \,  h^{-1} }
\newcommand{\hOMpc}{ \,  \mathrm{Mpc}^{-1}   h  }

\newcommand{\comment}[1]{}

\newcommand{\dv}{ \delta_{\rm v}}

\newcommand{\Msun}{ \,   M_{\odot}  {h}^{-1}   }

\newcommand{\beq}{\begin{equation}}
\newcommand{\eeq}{\end{equation}}
\newcommand{\beqa}{\begin{eqnarray}}
\newcommand{\eeqa}{\end{eqnarray}}

\newcommand{\revise}[1]{\textcolor{black}{#1}}

\begin{document}

\title{ Measurement of Void Bias Using Separate Universe Simulations }

\correspondingauthor{Kwan Chuen Chan}
\email{chankc@mail.sysu.edu.cn}

\author{Kwan Chuen Chan}
\affiliation{ School of Physics and Astronomy, Sun Yat-Sen University, Guangzhou 510275, China }


\author{Yin Li}
\affiliation{ Kavli Institute for the Physics and Mathematics of the Universe (WPI), UTIAS, The University of Tokyo, Chiba 277--8583, Japan }
\affiliation{ Berkeley Center for Cosmological Physics, Department of Physics, \\ \& Lawrence Berkeley National Laboratory,  University of California, Berkeley, CA 94720, USA }

\author{Matteo Biagetti}
\affiliation{ Institute of Physics, Universiteit van Amsterdam, Science Park 904, 1098XH Amsterdam, the Netherlands   }

\author{Nico Hamaus}
\affiliation{  Universit\"ats-Sternwarte M\"unchen, Fakult\"at f\"ur Physik, Ludwig-Maximilians Universit\"at, Scheinerstr.~1, 81679 M\"unchen, Germany}




\begin{abstract}

Cosmic voids are biased tracers of the large-scale structure of the universe.  Separate universe simulations (SUS) enable accurate measurements of this biasing relation by implementing the peak-background split (PBS).  In this work, we apply the SUS technique to measure the void bias parameters.  We confirm that the PBS argument works well for underdense tracers.  The response of the void size distribution depends on the void radius.  For voids larger (smaller) than the size at the peak of the distribution, the void abundance responds negatively (positively) to a long wavelength mode.  The linear bias from the SUS is in good agreement with the cross power spectrum measurement on large scales.
Using the SUS, we have detected the quadratic void bias for the first time in simulations.  We find that $ b_2 $ is negative when  the magnitude of $ b_1  $ is small, and that it becomes positive and increases rapidly  when  $ |b_1|  $ increases. We compare the results from voids identified in the halo density field with those from the dark matter distribution, and find that the results are qualitatively similar, but the biases generally shift to the larger voids  sizes.

\end{abstract}


\keywords{ (cosmology:) large-scale structure of universe}


\section{ Introduction}
Cosmic voids trace the underdense regions of the large-scale structure.  Similar to halos, they are biased with respect to the dark matter density field \citep{Sheth:2003py}.   They hold information  complementary to that of the halos.  Voids are generally large in size, so a big survey volume is required to obtain  good statistics on voids. As galaxy surveys explore larger and larger volumes, we anticipate that voids as a large-scale structure tracer will play a more prominent role in the future.

Void bias has been studied in simulations \citep{Hamaus:2013qja,Chan:2014qka,Chan:2018piq,SchusterHamaus_etal2019}, and it has been measured using SDSS data \citep{Clampitt:2015jra}. These analyses rely primarily on the measurements of two-point functions: power spectrum in Fourier space or correlation function in configuration space. It was found that the large-scale void bias can be either positive or negative, depending on the void size and redshift.  In contrast, the halo bias is always positive. Although often regarded as a nuisance parameter, \citet{Chan:2018piq} showed that the void bias parameter exhibits large-scale scale dependence in the primordial non-Gaussian (PNG) model analogous to the well-known PNG halo bias (\citet{Dalal:2007cu}; see \citet{Biagetti2019} for a review), and this can significantly tighten the bound on the PNG parameter in upcoming surveys.  \revise{ Beyond the linear biases, we anticipate the existence of a local quadratic void bias $b_2 $. The nonlinear bias parameter reflects the nonlinear response of the void abundance to a long wavelength perturbation and is important for the understanding of void formation physics. One can also measure the clustering of voids using three-point statistics such as the (cross-) bispectrum. In this case, at tree level besides the local quadratic bias $b_2$, one would also have the non-local bias \citep{McDonaldRoy,Chan:2012jj,Baldaufetal2012}. The three-point statistics of voids could help to break the degeneracy between $\sigma_8$ and $b_1$, as in the case of halos. }

Another approach to study the bias parameter is to use the separate universe simulations (SUS; \citet{McDonald2003,Sirko2005}).   The idea of the SUS is to absorb the local long wavelength perturbation into the background of a separate universe.  By so doing, the effect of the long mode can be taken into account by simply changing the parameters of the simulations without modifying the code. \citet{LiHuTakada2014} developed paired SUS to measure the response of the nonlinear power spectrum to the local background density $\delta_\mathrm{b}$.  Later, they have been generalized nonlinearly in $\delta_\mathrm{b}$ in \citet{Wagner_etal2015}.   The SUS furnish an accurate implementation of the peak-background split (PBS; \citet{Kaiser1984,Mo:1995cs,ShethTormen1999,Desjacques:2016bnm})  in the long wavelength limit.  They  enabled the precise measurement of the local halo bias parameters  \citep{LiHuTakada2016, Lazeyras_etal2016,BaldaufSeljakSenatoreZaldarriaga2016} and test of the assembly bias \citep{Paranjape:2016pbh,Lazeyras:2016xfh}.
Besides, the SUS have been used to study the degeneracy between the survey long wavelength modes and cosmological parameters \citep{LiHuTakada2014sss, LiSchmittfullSeljak2018}; the squeezed $N$-point function \citep{ChiangWagner_etal2014, WagnerSchmidt_etal2015JCAP}; the baryonic effects on the matter power spectrum \citep{Barreira_etal2019},  the effects of other smooth components, such as quintessence dark energy or neutrinos on the power spectrum and halo abundance \citep{HuChiangLiLoVerde2016, ChiangLiHuLoVerde2016, ChiangHuLiLoVerde2018}; and the Lyman-$\alpha$ forest \citep{McDonald2003,CieplakSlosar2016}.

Moreover, \citet{Chan:2018piq} found that the PNG void bias parameter as a response of the void abundance to the local value of $\sigma_8$ yields good agreement with the numerical results. That prediction was obtained by finite differencing the void size distribution from simulations with different values of $\sigma_8$.  Because this procedure is very similar to the SUS in spirit, this success motivates us to apply the SUS to measure the void bias. This is the first time that the SUS have been applied to study  underdense tracers, and we shall see that they enable us to directly test the PBS argument on underdense tracers.  This paper is organized as follows. We first review the SUS technique in Sec.~\ref{sec:SUS_review} and then provide the details of the simulations used in this work in Sec.~\ref{sec:simulations_details}. The measurements of the void bias from dark matter voids and halo voids are presented in Sec.~\ref{sec:void_bias_measurement}.  We conclude in Sec.~\ref{sec:conclusion}.  Appendix \ref{sec:LagEul_mapping} is devoted to the discussion on the mapping of the void size distribution from Lagrangian to  Eulerian space.  We discuss the generalization of the abundance matching method to the second order in Appendix \ref{sec:ABmatching}.

\section{Review of the SUS technique}
\label{sec:SUS_review}

The principal idea of the SUS is that locally a constant long wavelength perturbation in the global universe can be absorbed into the background of a separate universe. The SUS technique enables the simulation of the effect of the long mode on small scales by simply changing the parameters of a separate simulation without the need to modify the code.  This method was introduced in \citet{McDonald2003,Sirko2005}. \citet{Wagner_etal2015} generalized it to the exact nonlinear order in the long mode and \citet{LiHuTakada2014} developed paired SUS techniques that enable calibration of observable responses without sample variance. In this work, we implement the nonlinear paired SUS to increase the signal to noise in void bias measurements.  Here, we review the essential ingredients of SUS following \citet{Sirko2005}, \citet{LiHuTakada2014}, and \citet{Wagner_etal2015}.

In a local patch of the universe, the mean overdensity $\delta_{\rm b} $ fluctuates on top of the global background.  Note that $\delta_\mathrm{b}$ can be nonlinear, and it can be absorbed into the background of a local universe as
\begin{align}
  \label{eq:matter_equality_universes_1}
\bar{\rho}_{\rm m}(t) [  1+\delta_{\rm b}(t) ] &= \bar{\rho}_{\rm m}^{W}(t) ,
\end{align}
where   $ \bar{\rho}_{\rm m} $  and $ \bar{\rho}_{\rm m}^{W} $ are the matter density in the global and local universes.  From here on, we use the script $W$ \revise{(for windowed)} to denote a quantity in the local universe and the quantities without it are in the global universe.   In terms of the scale factors $a$ and $a_W$, we can write Eq.~\eqref{eq:matter_equality_universes_1} as
\begin{align}
\label{eq:matter_equality_universes_2}
\frac{H_0^2  [ 1 + \delta_{\rm b}(t) ] \Omega_{\rm m} }{ a^3(t)  } & = \frac{ H_{0W}^{2 }\Omega_{\rm m}^W }{ a_W^3(t) },
\end{align}
where $H_0$ and $\Omega_{\rm m}$ ($H_{0W}$ and $\Omega_{\rm m}^W$) are the Hubble parameter and the matter density parameter in the global (local) universe today.

For the global universe, it is convenient to use the standard convention $a=1$ at the present time,  while for the separate universe we use the convention $a_W \rightarrow a$ as  $a\rightarrow 0 $.
Because $\delta_{\rm b} \rightarrow 0$ as $a\rightarrow 0 $, we deduce that
\begin{equation}
  H_{0}^{2} \Omega_{\rm m} = H_{0 W}^{2} \Omega_{\rm m}^{W}.
\end{equation}
Hence Eq.~\eqref{eq:matter_equality_universes_2} implies that
\begin{align}
 \label{eq:a_separate_universe_2}
a_{W}(t) &=  [ 1+ \delta_{\rm b}(t) ]^{-\frac{1}{3}} a(t) .
\end{align}

We take the global universe to be a flat $\Lambda$CDM, and as a result of $\delta_{\rm b} $ in the global universe, the local one acquires a curvature  $K_W$
\begin{align}
  \label{eq:KW_exact}
  \frac{K_{W}}{ H_0^{2}} & 
    =  \frac{5}{3} \frac{\Omega_{\rm m} \delta_{\rm b0}^{\rm L} }{D_{0}},
\end{align}
where $D_0$ is the present-day linear growth factor reducing to the scale factor in the matter-dominated regime, and $ \delta_{\rm b 0}^{\rm L} $ is the perturbation linearly extrapolated to the present epoch. Because $K_{W}$ is a conserved quantity, it can be evaluated at the early epoch when linear theory suffices. Then  $H_W $ follows from
\begin{equation}
  \label{eq:deltaH_exact}
\frac{ H_{0W}  }{ H_0 } \equiv  1+\delta_{H}=\sqrt{ 1-\frac{K_{W}}{H_{0}^{2}} }.
\end{equation}
The density parameters can be obtained by the relations
\begin{align}
  \Omega_{\rm m}^W &= \frac{\Omega_{\rm m}}{\left(1+\delta_{H}\right)^{2}} , \\
  \Omega_{\Lambda}^W &= \frac{\Omega_{\Lambda}}{\left(1+\delta_{H}\right)^{2}}, \\
  \label{eq:Omega_k_deltah}
  \Omega_{K}^W &= 1 -  \frac{1}{\left(1+\delta_{H}\right)^{2}}.
\end{align}

To match the results in the local universe with the global one, we need to output the simulations at the same physical time, $t$.  We can match $a_W $ with the fiducial $a $ by demanding
\begin{equation}
  \label{eq:t_match_condition}
t = \int_{0}^{a} \frac{d a'}{a' H(a')}=\int_{0}^{a_{W} } \frac{d  a_W' }{ a_W' H_W(a_W')}.
\end{equation}

In summary, a local patch evolves as a separate universe with curved
$\Lambda$CDM cosmology, related to the global one by the above equations.

\section{Details of the simulations}
\label{sec:simulations_details}
The cosmological parameters of the fiducial simulations are  $\Omega_{\rm m}=0.3  $,  $\Omega_{\Lambda}=0.7  $, $h_0 =0.7$, and  $\sigma_8=0.85 $. Each simulation has $ 512^3 $ particles.  Two sets of SUS with different box sizes were run. The fiducial box sizes are  $666.667 \MpcOh$ (medium-size set, M-set) and  $333.333 \MpcOh$ (high-resolution set, H-set), respectively. For the M-set, we ran SUS with $\delta_{\rm b0}^{\rm L} = 0$,  $\pm0.05$, and $\pm 0.2$, while we have $\delta_{\rm b0}^{\rm L} = 0$,  $\pm0.1$, and $\pm 0.3$  for the H-set. For both sets, there are six realizations for each $\delta_{\rm b0}^{\rm L} $. For convenience, we have listed some of the information of the simulations in Table \ref{tab:sim_info}.

The Gaussian initial conditions are generated by {\tt CLASS} \citep{CLASS_code}.  We use the linear growth factor scaling described in \citet{Wagner_etal2015} to generate the power spectrum of the SUS from the fiducial one. The particle displacements are implemented using {\tt 2LPTic}~\citep{CroccePeublasetal2006} at $z=49$, and  evolved with the $N$-body code {\tt Gadget2} \citep{Gadget2}.

The box size of the local universe $L_W$ is determined by matching the comoving size with the fiducial value $L$ as \citep{LiHuTakada2014}
\beq
\frac{ L }{ h} = \frac{ L_W }{ h_W }.
\eeq
We have made the Hubble parameter in the unit explicit. In the comoving matching scheme, the particle mass is the same irrespective of $\delta_{\rm b}$.

We need halo samples as we consider voids constructed on the halo density field.  Halos are identified using the spherical overdensity halo finder {\tt AHF}~\citep{AHF_2009}. They are defined with the spherical overdensity threshold of 200 times of the background density in the fiducial cosmology.  We use halos with at least 20 particles.  In order to get a sample with the same physical properties, we adjust the threshold for the separate universe as \citep{LiHuTakada2016, Lazeyras_etal2016}
\begin{align}
\rho_{\rm th}^{\rm h} =  200 \rho_{\rm m} ( t )
= \frac{ 200 }{ 1 + \delta_{\rm b} (t)  } \rho_{\rm m}^W(t).
\end{align}
As \citet{Lazeyras_etal2016}, we turn off the option of removing unbound particles in  {\tt AHF}.


\revise{ There are numerous void finders in the literature, see e.g.~\citet{Colbergetal2008}; here, we use the void finder {\tt VIDE}~\citep{Sutter:2014haa}}, which is based on {\tt ZOBOV}~\citep{Neyrinck2008} using a watershed algorithm~\citep{Platenetal2007}. In this work, we consider voids created out of the dark matter and halo density field in real space.  In {\tt VIDE}, after the construction of local basins (zones) via  the watershed algorithm, zones are expanded until some threshold linking-density between them is reached.  The default threshold value is 0.2 of the background density.  Similar to the halo case, to ensure that the expansion is stopped at the same physical density, we rescale the density threshold as
\begin{equation}
    \rho_\mathrm{th}^\mathrm{v} = 0.2 \rho_\mathrm{m}(t)
    = \frac{ 0.2 }{ 1 + \delta_{\rm b} (t) } \rho_\mathrm{m}^W(t).
\end{equation}
If this expansion feature is turned off and the zones from the watershed algorithm are taken to be voids directly, the difference on the void abundance is only noticeable for large voids, and it is negligible compared with the statistical fluctuations.

Voids generally depend on the resolution and when the tracer density increases more and more sub-structures will be resolved.  For the identification of voids in matter distribution with {\tt VIDE}, we need to specify the subsampling density, the number density of the randomly selected dark matter tracer particles. We always quote the absolute subsample density in the fiducial simulation ($\delta_{\rm b 0}^{\rm L} = 0$).  For the SUS, the {\it relative} subsampling density is set to match that of the fiducial one, i.e., the downsampling fractions are the same for the SUS. For the halo voids, we construct voids on the halo sample with the same minimum halo mass.

As voids are extended objects, we need to match their physical size to compare the voids identified across different SUS with
\beq
\label{eq:phy_void_size_matching}
a R = a_W R_W .
\eeq
In practice, we first change the unit of the void size from the SUS to  $ \mathrm{ Mpc}  h^{-1} $   by multiplying by a factor of  $h / h_W $. Then we convert the void radius from the SUS to the comoving size in the fiducial cosmology by multiplying a factor of $ a_W / a $.  In fact, this is analogous to the case of power spectrum comparison, in which one matches the physical wavenumber with the condition $ k/a = k_W/ a_W $ \citep{LiHuTakada2014}.

In addition, we compare the SUS results against the cross power spectrum measurement from simulations of large box size (denoted as L-set). These simulations are the Gaussian runs in \citet{Biagetti:2016ywx} and the void bias measurements have been presented in \citet{Chan:2018piq}.  Their cosmology is the same as the fiducial values of the SUS. In each simulation, there are $1536^3$ particles in a cubic box of side length 2000 $\MpcOh$. In fact, the resolution of the M-set is chosen to match that of the L-set.  There are eight realizations in total.

\begin{table}[t!]
\centering
 \begin{tabular}{*5c}
 \toprule
 Set & $ L_{\rm box}/ \,  \MpcOh$  & $N_{\rm particles} $   & $\delta_{\rm b 0}^{\rm L}$ & Realizations  \\ [0.5ex]
 \midrule
 M & 666.667 & $512^3$ & $0, \, \pm0.05, \, \pm0.2$ & $5 \times 6$ \\
 H & 333.333 & $512^3$ & $0, \, \pm0.1, \, \pm0.3$ & $5 \times 6$ \\
 L & 2000 & $1536^3$ & 0 & 8 \\
 \bottomrule
 \end{tabular}
 \caption{ Three sets of simulations used in this work.   }
\label{tab:sim_info}
\end{table}

\begin{figure*}[!htb]
\centering
\includegraphics[width=0.9\linewidth]{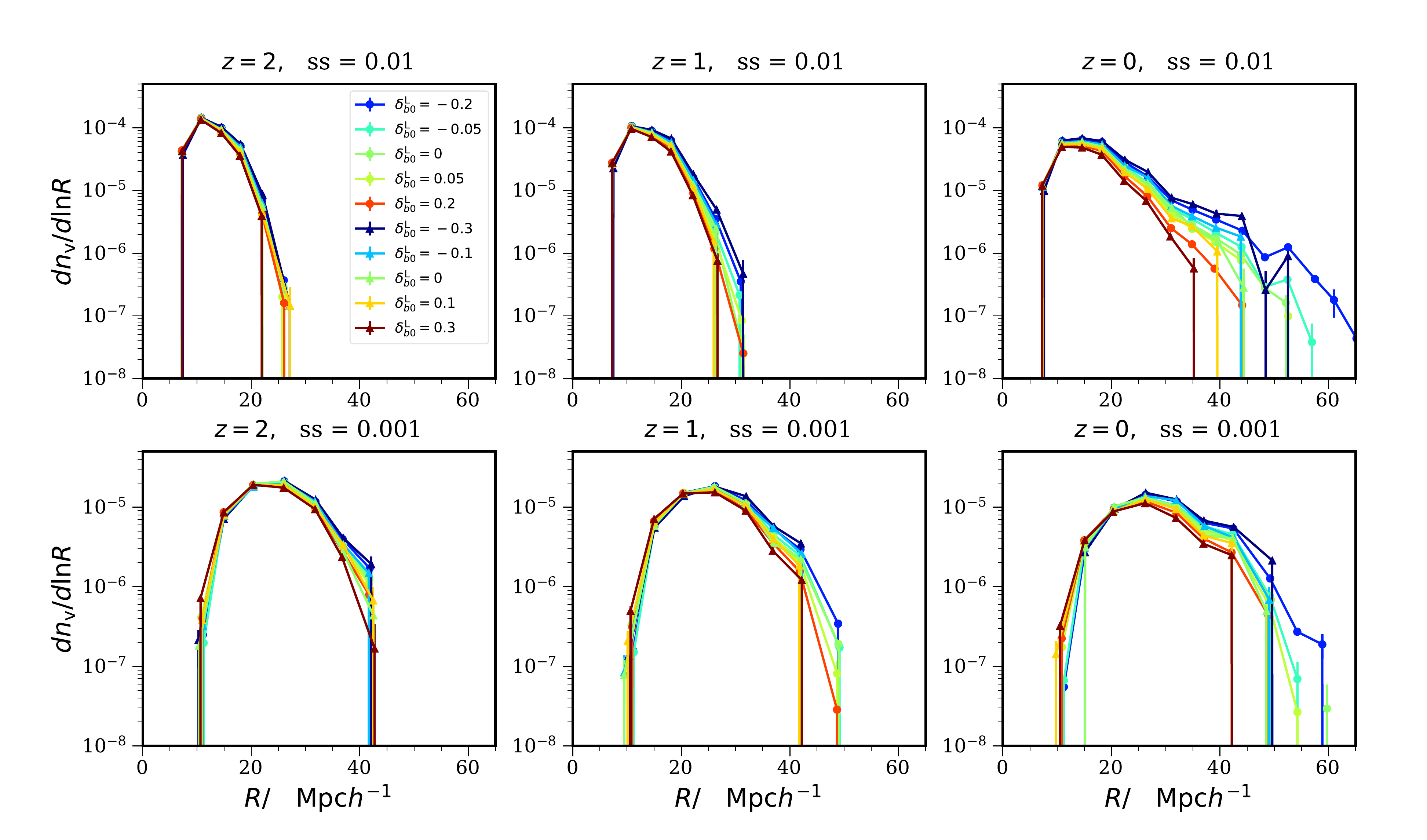}
\caption{  Void size distribution for different values of $\delta_{\rm b0}^{\rm L}$: $\pm 0.2$, $\pm 0.05 $, and 0 from the M-set (circles) and  $\pm 0.3$, $\pm 0.1 $, and 0 from the H-set (triangles). The voids are constructed from the dark matter density field at $z=2$, 1, and 0 (left to right panels). The results obtained using the subsampling density  0.01 and 0.001 $(\MpcOh)^{-3}$  are compared (top and bottom panels).  }
\label{fig:dnlnR_DMvoids_MHsets_subset}
\end{figure*}

\begin{figure}[!htb]
\centering
\includegraphics[width=\linewidth]{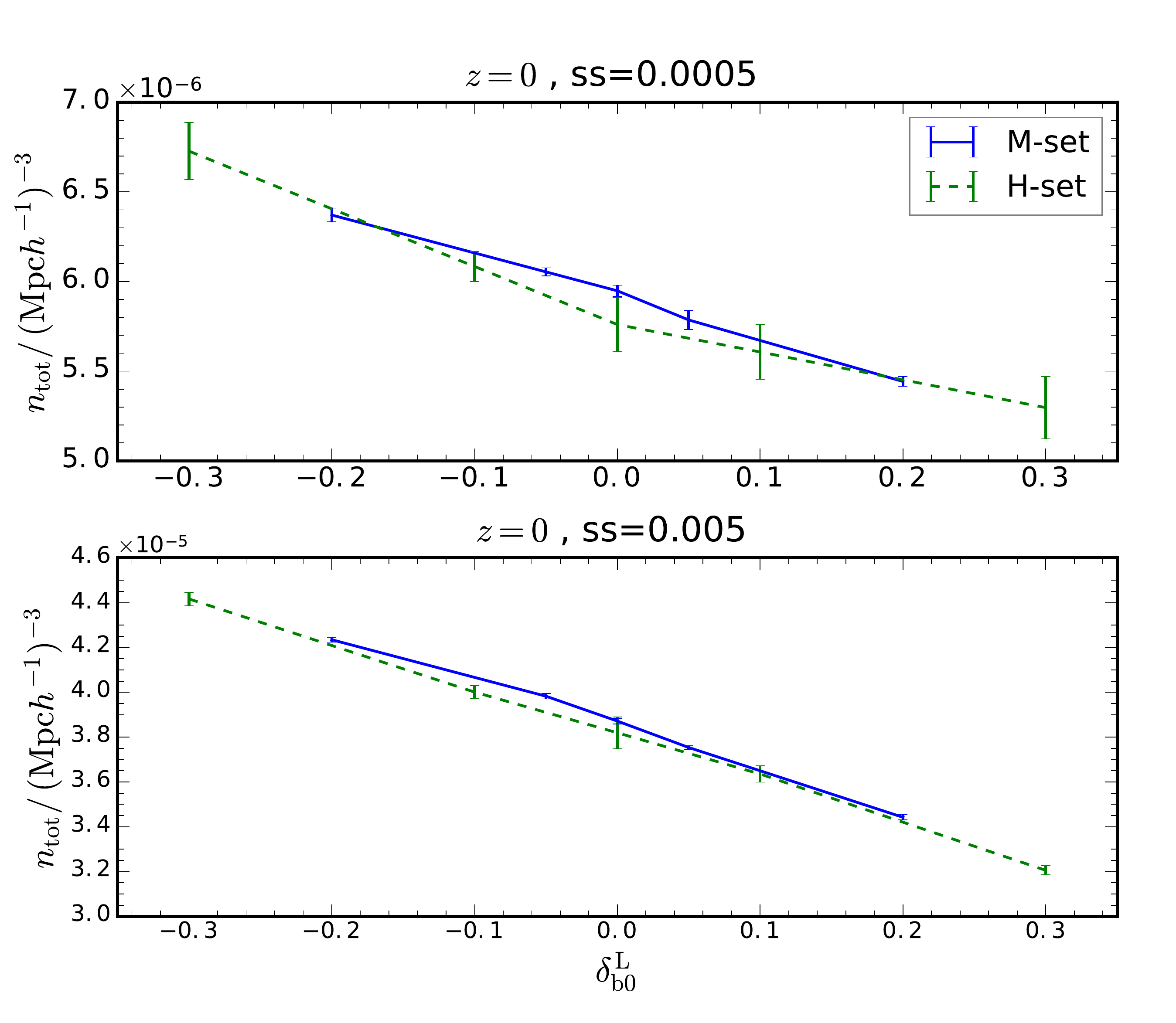}
\caption{  Number density of all the voids as a function of $\delta_{\rm b0}^{\rm L} $. The results from M-set (solid, blue) and H-set (dashed, green) at $z=0$ obtained with subsampling density 0.0005 and 0.005 $(\MpcOh)^{-3}  $ are shown. }
\label{fig:ntot_deltab}
\end{figure}

\begin{figure*}[!htb]
\centering
\includegraphics[width=0.9\linewidth]{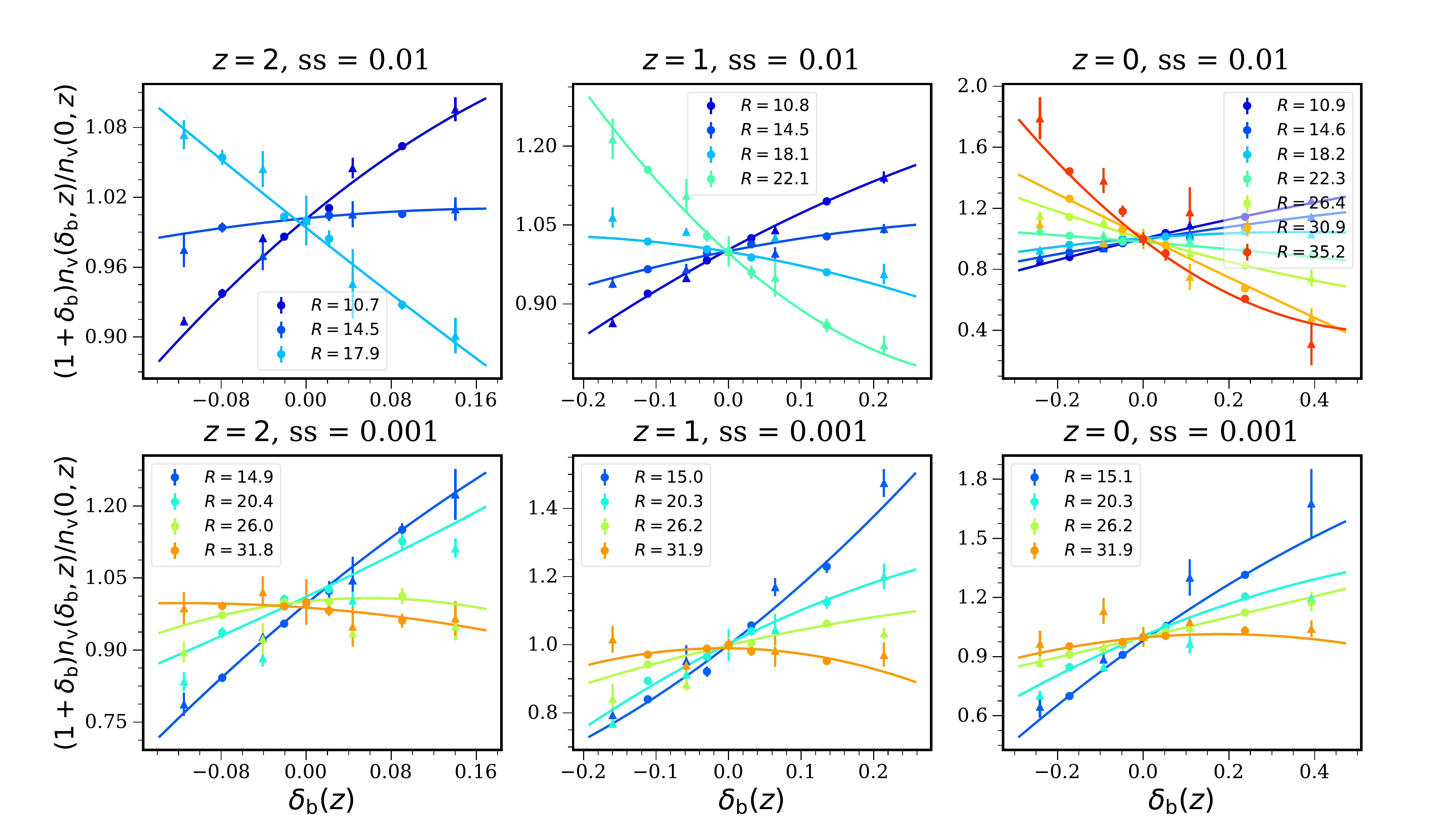}
\caption{ Normalized void size distributions multiplied by $(1 + \delta_{\rm b} )$  as a function of $\delta_{\rm b}$.  The results [M-set (circles) and H-set (triangles)] obtained with various subsampling densities [0.01 and 0.001 $(\MpcOh)^{-3}$)] and  at different redshifts (2, 1, and 0) are shown. The mean void sizes \revise{(in units of $\MpcOh$)} are shown in the legend and the best-fit  quadratic polynomial is also plotted.   }
\label{fig:nv_DMvoids_deltab_Eul_subset}
\end{figure*}

\begin{figure*}[!htb]
\centering
\includegraphics[width=0.9\linewidth]{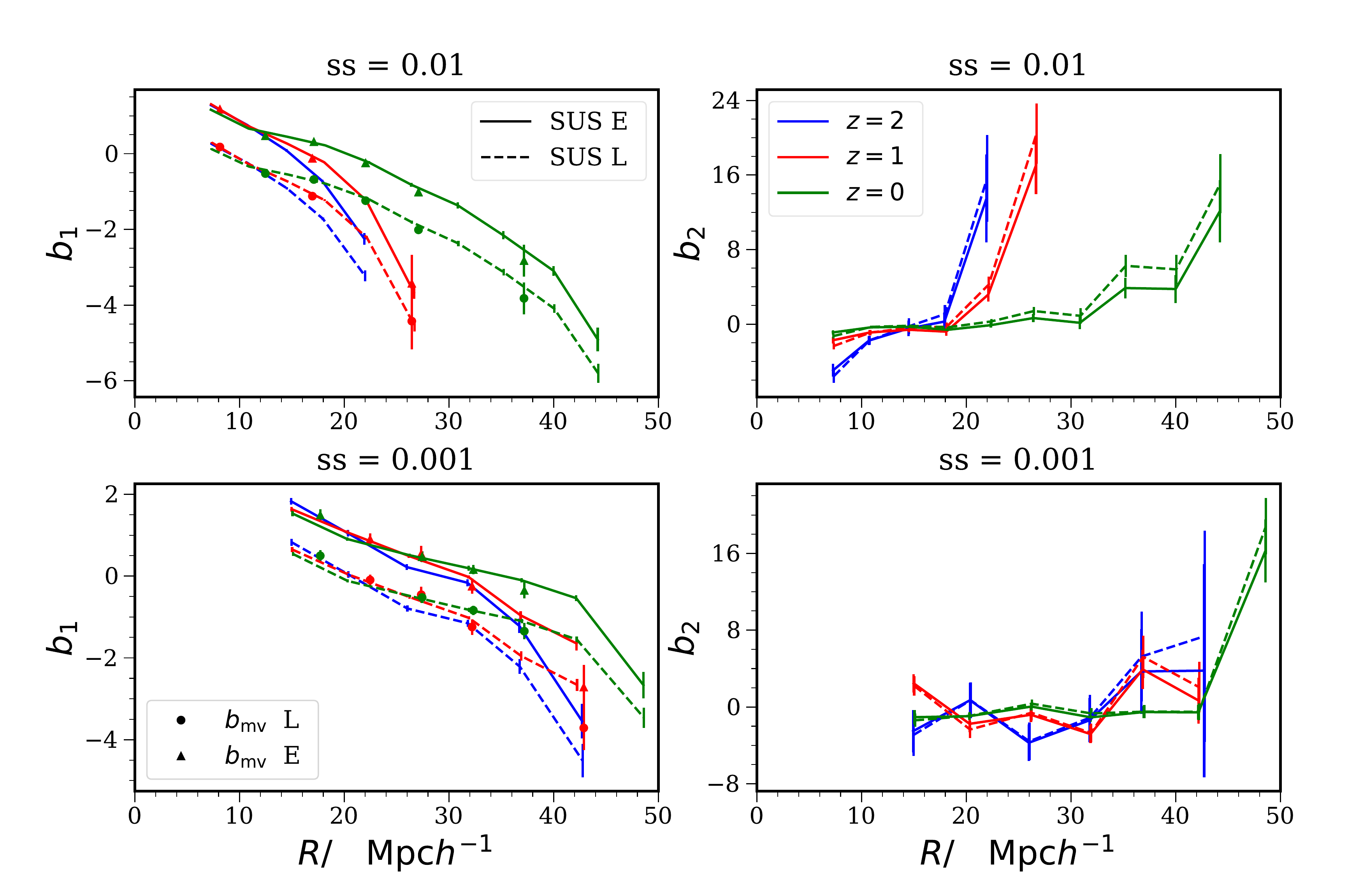}
\caption{  Linear and quadratic SUS void bias as a function of void radius (left and right panels, respectively). The Lagrangian (dashed lines) and Eulerian  (solid lines) results at $z = 2$ (blue), 1 (red), and 0 (green) are shown. The Eulerian linear bias $b_{\rm mv}$ from the large-scale cross power spectrum measurement (triangles) and the derived Lagrangian results (circles) are shown for comparison.   }
\label{fig:void_bias_DMvoid_subset}
\end{figure*}

\begin{figure}[!htb]
\centering
\includegraphics[width=\linewidth]{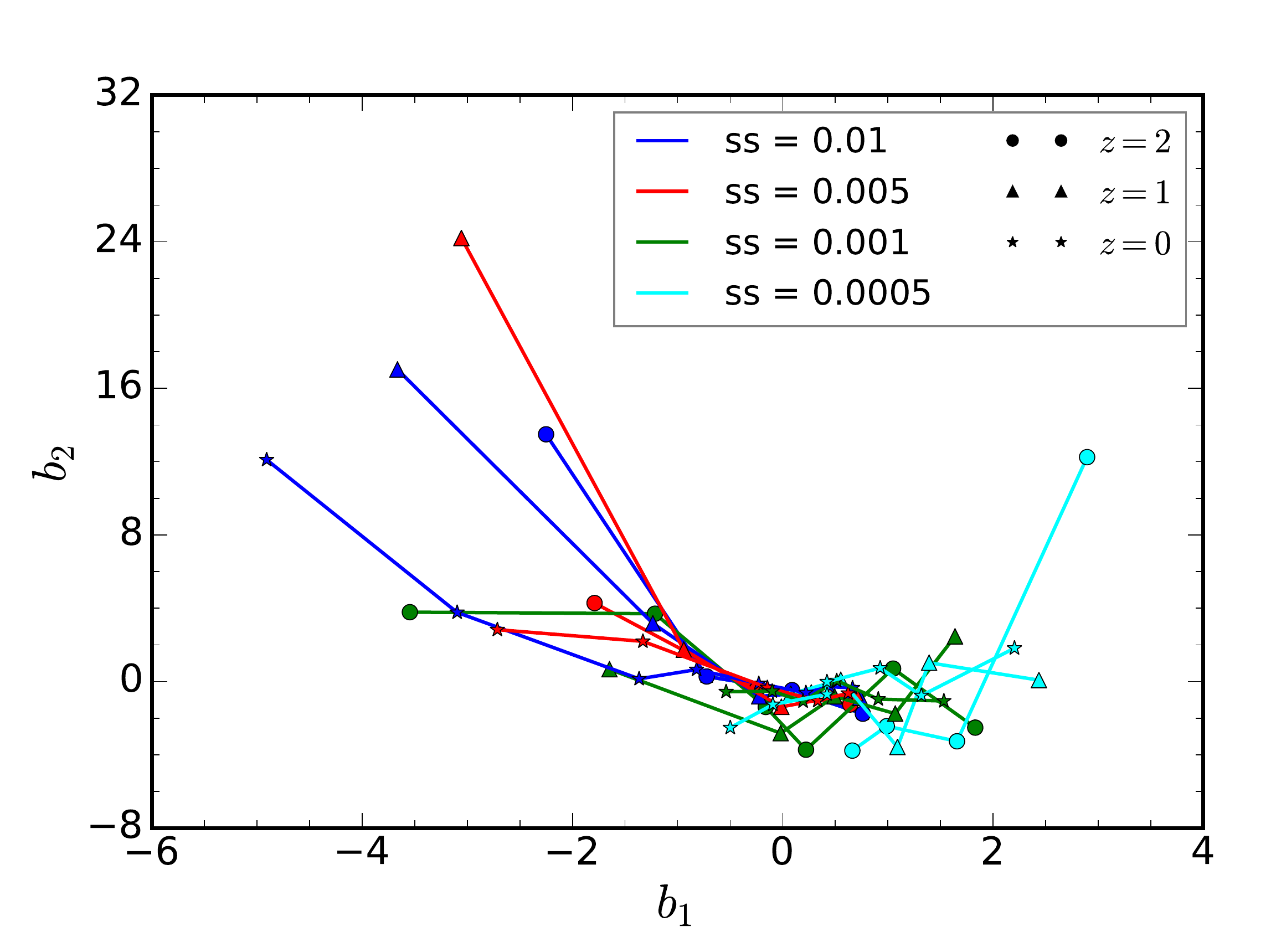}
\caption{  Best-fit SUS $b_2 $ versus $b_1$. The results are derived from the dark matter void samples at $z=2$ (circles), 1 (triangles), and 0 (stars) with subsampling density of 0.01 (blue), 0.005 (red), 0.001 (green), and 0.005 (cyan) $(\MpcOh)^{-3} $, respectively.  }
\label{fig:b1b2_DMvoids}
\end{figure}

\section{ Void bias measurements}
\label{sec:void_bias_measurement}

We show the void bias measurements using the SUS in this section.  We first consider the dark matter void results and then move to the halo voids. The default method is analogous to the SUS halo bias measurement presented in \citet{Lazeyras_etal2016}. We first measure the void size distribution from the SUS, and then fit a quadratic polynomial in  $\delta_{\rm b} $  to the resultant void size function to get the PBS bias parameters. We review the theoretical background of this method in Appendix \ref{sec:LagEul_mapping}.  See the Appendix \ref{sec:ABmatching} for an alternative abundance matching method.

\subsection{Dark matter void bias}

\revise{ \subsubsection{Response of the void size distribution} }

We first consider voids identified in the dark matter density field. In Fig.~\ref{fig:dnlnR_DMvoids_MHsets_subset}, we show the void size distribution measured from the SUS with different values of $\delta_{\rm b}^{\rm L} $. The results at $z=2$, 1, and 0 are presented. Two subsampling densities corresponding to 0.01 and 0.001  $(\MpcOh)^{-3}$ in the fiducial setting are shown. All of the SUS have the same relative subsampling density as the fiducial one.  \revise{ Unless otherwise stated, the error bars are estimated by the standard deviations of the measurements among the six realizations. }

We find that the void abundance above and below the void size corresponding to the peak of the distribution (peak size) responds  differently to $\delta_{\rm b}^{\rm L}  $. For convenience, let us call the region with void sizes smaller than the peak size, the {\it positive response} (PR) regime, and the one above it the {\it negative response} (NR) regime.

For voids in the NR region, as $\delta_{\rm b}^{\rm L} $ increases the void abundance reduces.  This is particularly obvious at the large-size end of the distribution. This trend is the opposite of the halos, for which  positive   $\delta_{\rm b}^{\rm L} $ enhances the halo abundance (but see below).  Physically these voids correspond to a large underdense region, when $\delta_{\rm b}^{\rm L} $ increases, the mean overdensity in this region increases and it becomes harder to form a void.

For the voids in the PR regime, we find that the void abundance actually increases with  $\delta_{\rm b}^{\rm L}  $. When the subsampling density is small [e.g.~0.001 $(\MpcOh)^{-3}$], the small-size end rises less abruptly and this trend is more apparent. Although it may sound counter-intuitive, this kind of response is also found in the halo mass function.  Using the H-set, we find that at $z=0$, for $ M \lesssim 3\times 10^{12} \Msun $, the halo abundance response flips sign and the abundance decreases with positive $\delta_{\rm b}  $.  This behavior can be understood using the halo model \citep{Cooray:2002dia}, in which all the matter is assumed to reside in halos of various mass. In the halo model, there is a consistency relation \citep{Cooray:2002dia}:
\beq
\label{eq:HM_integral_CR}
\int d \ln M \frac{M}{\bar{\rho}} \frac{dn}{d \ln M } b_1 = 1 ,
\eeq
and the integral vanishes for higher-order biases.  As the massive halos have bias greater than 1, the low-mass halos must have bias less than 1.  The substance of Eq.~\eqref{eq:HM_integral_CR} is the conservation of mass. Physically, when positive $\delta_{\rm b} $ promotes the formation of massive halos, they suck up so much mass that there is little mass available for the small halos to form.  The value $ 3\times 10^{12} \Msun $ corresponds to  the mass at which the Lagrangian bias switches sign at $z=0$.  We note that the simple bias models \citet{Mo:1995cs} and \citet{Scoccimarroetal2001} both predict that the Lagrangian bias switches sign at  $M_* $, the characteristic mass scale of the halo mass function. At $z=0$, for our fiducial cosmology it is about  $ 5.6 \times 10^{12} \Msun $. It is slightly different from the value we found. However, this is not very surprising  because these excursion set theories fail to work well in the low-mass end.

The analog void model for matter is less tenable, but at least it may give some insights into why the void abundance response switches sign.  A possible physical scenario is that when $\delta_{\rm b}^{\rm L} $ increases, the previously smooth region collapses and splits up into a few domains, and this generates  shallow regions in between them. Because the shallow regions are surrounded by the overdense regions, they are not as underdense as the voids in the NR regime.  These are the so-called over-compensated voids with high ridges and like halos matter falls onto them \citep{HamausSutterWandelt2014}.  As we see below, they tend to have sizable positive linear bias.  This regime is difficult to explain using the excursion set theory. E.g.,~the void-in-cloud scenario \citep{Sheth:2003py} alone would reduce the void abundance rather than increase it because positive  $\delta_{\rm b}^{\rm L} $ makes it harder to form voids and easier for the large scale to collapse. Nonetheless, the physical picture is consistent with the low-mass end of the halo mass function. Both are consequences of mass  conservation. The shallow regions appear because the collapsed structures suck up the mass in the environment.

So far, we have looked at the void abundance at fixed $R$ and discuss how it responds to $\delta_{\rm b}$.  There is an alternative viewpoint from the abundance matching perspective [see \citet{LiHuTakada2016} and Appendix \ref{sec:ABmatching}]. In this interpretation, it is assumed that voids from different SUS respond to $\delta_{\rm b} $ by shifting in size.  For positive $ \delta_{\rm b}$, an underdense region will destine to form void of smaller size.  For the range of  $\delta_{\rm b}$ we consider, a significant fraction of the voids respond to $\delta_{\rm b}$ by going through this path.  For many others, however, increase of  $\delta_\mathrm{b}$ make the mean density higher than the density threshold and prevent void formation. Increase of $\delta_{\rm b}$  can also cause segmentation  as explained in the last paragraph.   By inspecting Fig.~\ref{fig:dnlnR_DMvoids_MHsets_subset}, we see that  the area decreased in the NR regime is larger than the area increased in the PR regime. In Fig.~\ref{fig:ntot_deltab}, we show the number density of all the voids against  $\delta_{\rm b } $.  We find that the number density decreases with $\delta_{\rm b } $, and so increase of  $ \delta_{\rm b}$ indeed causes some region to fail to form voids.

\revise{ The resolution dependence of the void size distribution can be seen in, e.g., Fig.~1 of \citet{Chan:2014qka}, Fig.~6 of \citet{Jennings:2013nsa}, and Fig.~2 of \citet{Ronconi:2019xex}.  When the number density of the tracer increases, more sub-structures are resolved and so the peak is shifted to a smaller scale.  Close inspection of these figures in the references also reveals that the void abundance from the high tracer-density sample is slightly lower than in the lower resolution one at large void sizes. Physically, this reduction is due to a fragmentation of large voids into smaller ones. Because of this resolution issue,  the agreement among the simulations with different resolutions and with the theory only holds for a limited range in size for a given resolution.  For the clustering analysis, we are more interested in how well voids trace the underlying large-scale structure. To maximize the information, we use all the available voids even though there is resolution dependence.   
}

\revise{ \subsubsection{Measurement of the bias from the response} }

\revise{ After obtaining the response of the void size distribution to the long mode, we are now in a position to extract the bias parameters.  }   The Eulerian PBS bias parameter can be obtained as (\citet{Mo:1995cs}; see also Appendix~\ref{sec:LagEul_mapping} for a review)
\beq
\label{eq:bi_Eulerian}
b_i = \frac{ \partial^i }{ \partial \delta^i_{\rm b} } \left[  (1 + \delta_{\rm b} ) \frac{ n_{\rm L}( \delta_{\rm b} ) }{ \bar{n}  } \right] \bigg|_{\delta_{\rm b }= 0  },
\eeq
where $  n_{\rm L}( \delta_{\rm b} ) $ and  $ \bar{n} $  denote the void size distribution with the long mode and without it.   In essence, the Eulerian bias parameters are the Taylor expansion coefficients of the fluctuations of the void size distribution in Eulerian space.  Hence we can extract the PBS bias parameters from the function $   (1 + \delta_{\rm b} ) [ n_{\rm L}( \delta_{\rm b} ) /  \bar{n} ] $ by fitting a quadratic polynomial  in $ \delta_{\rm b}(z)$ to it.  We have plotted the results in Fig.~\ref{fig:nv_DMvoids_deltab_Eul_subset} for a number of void size bins (bin width of $5 \MpcOh$). \revise{The best fit and its error are obtained using the {\tt numpy} least-square fitting routine {\tt polyfit} with the fit weighted inversely by the standard deviations among the realizations. }  We have combined the results from the M-set with the H-set, which has only 1/8 of the volume of the M-set and hence has less weight in the fitting.  We tried using a cubic polynomial to extract the bias parameters. The linear bias is not sensitive to it, but the quadratic bias becomes more noisy. Thus, to increase the signal to noise, in this work we stick to the quadratic bias model.

Let us also introduce the Lagrangian bias parameter, which is defined as
\beq
\label{eq:bi_Lagrangian}
b_i^{\rm L} = \frac{ \partial^i }{ \partial \delta_{\rm b}^{ {\rm L}i  } }  \left[  \frac{ n_{\rm L}( \delta_{\rm b}^{\rm L} ) }{ \bar{n}  } \right] \bigg|_{\delta_{\rm b  }^{\rm L}=0 },
\eeq
where  $ \delta_{\rm b }^{\rm L}  $ is the linearly extrapolated long wavelength perturbation. The Lagrangian and Eulerian PBS bias parameters can be related to each other using  spherical collapse via Eqs.~\eqref{eq:b1_MassConserve} and \eqref{eq:b2_MassConserve}. We obtain the Lagrangian bias following a procedure similar to the Eulerian one by fitting a quadratic polynomial in $ \delta_{\rm b }^{\rm L}  $ to  $ n_{\rm L}( \delta_{\rm b}^{\rm L} ) /  \bar{n} $.

The corresponding linear and quadratic bias parameters are displayed in Fig.~\ref{fig:void_bias_DMvoid_subset}. To verify the linear bias results, we have measured the large-scale bias using the cross power spectrum as \citet{Chan:2014qka}. The cross bias is defined as
\beq
b_{\rm mv} \equiv \frac{ P_{\rm mv}  }{ P_{\rm m} },
\eeq
where $ P_{\rm mv}$ is the cross power spectrum between voids and matter, and $ P_{\rm m} $ is the matter power spectrum.  $b_{\rm mv} $ is estimated using the $ \delta_{\rm b }^{\rm L} =0 $ suite of the M-set, and we have fitted a constant  up to $ k_{\rm max} = 0.04 \hOMpc $ to get the large-scale bias. The resultant bias is the Eulerian bias, and we derive the Lagrangian one using Eq.~\eqref{eq:b1_MassConserve}. We only use the  $b_{\rm mv} $ that is well fitted by a straight line (with  $\chi^2$ per degree of freedom less than 2). We will comment more on this in Sec.~\ref{sec:further_discussions}.  We find \revise{good} agreement between the SUS results and the cross power spectrum measurements.  In particular, we verify Eq.~\eqref{eq:b1_MassConserve} for voids (\citet{Massara:2018dqb} also arrived at a similar result).  The good agreement shows that the PBS argument works well for underdense tracers.   Our procedure mirrors the one for halos closely \citep{LiHuTakada2016,Lazeyras_etal2016}, and it shows that the modeling of the void size distribution and the void bias can be done in the same way as for the halos. We will discuss this further in Appendix~\ref{sec:LagEul_mapping}.

We also plot the best-fit quadratic void bias in Fig.~\ref{fig:void_bias_DMvoid_subset}. The signal-to-noise is highest for the high subsampling density and the low redshift samples. The overall trend that $b_2$ is a small negative number for small voids and then rapidly increases when the void radius increases  qualitatively  agrees with the theory prediction in \cite{Chan:2014qka}. We will comment more on this in Sec.~\ref{sec:further_discussions}. This is the first time that the quadratic void bias has been detected in simulations.

Finally, we plot the best-fit SUS $b_2$ against $b_1$ in Fig.~\ref{fig:b1b2_DMvoids}. We use data at $z=2$, 1, and 0  with the subsampling density 0.01, 0.005, 0.001, and 0.0005 $(\MpcOh)^{-3} $, respectively. There is no clear systematic trend with redshift, except perhaps the negative end of $b_1$, which corresponds to largest voids in the sample.  It is clear that as the subsampling density increases the bias parameters shift from the left part of the plot to the right.  In this plot, we see that towards negative $b_1 $, $b_2 $ goes up. The upturn of $b_2$ in the positive $b_1$ regime is less clear. It is interesting to note that when $b_1$ vanishes, $b_2 $ is close to zero.

\begin{figure*}[!htb]
\centering
\includegraphics[width=0.9\linewidth]{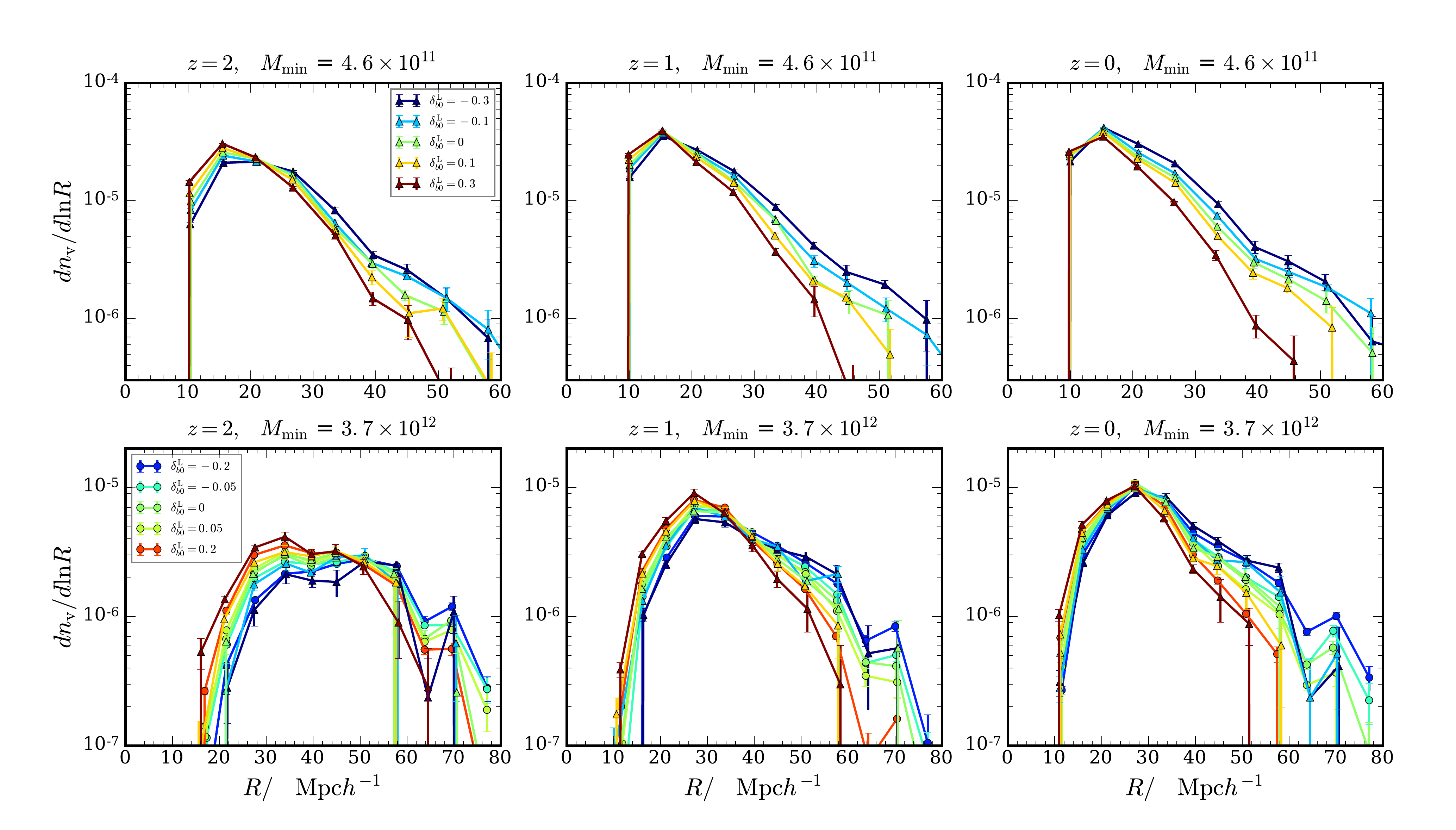}
\caption{  Similar to Fig.~\ref{fig:dnlnR_DMvoids_MHsets_subset}, but for halo voids obtained using the halo sample with minimum mass $4.6 \times 10^{11} $ (top panels, data from H-set only) and $3.7 \times 10^{12} \Msun$ (lower panels, data from both H-set and M-set). }
\label{fig:dnlnR_Halovoids_MHsets_subset}
\end{figure*}

\begin{figure*}[!htb]
\centering
\includegraphics[width=0.9\linewidth]{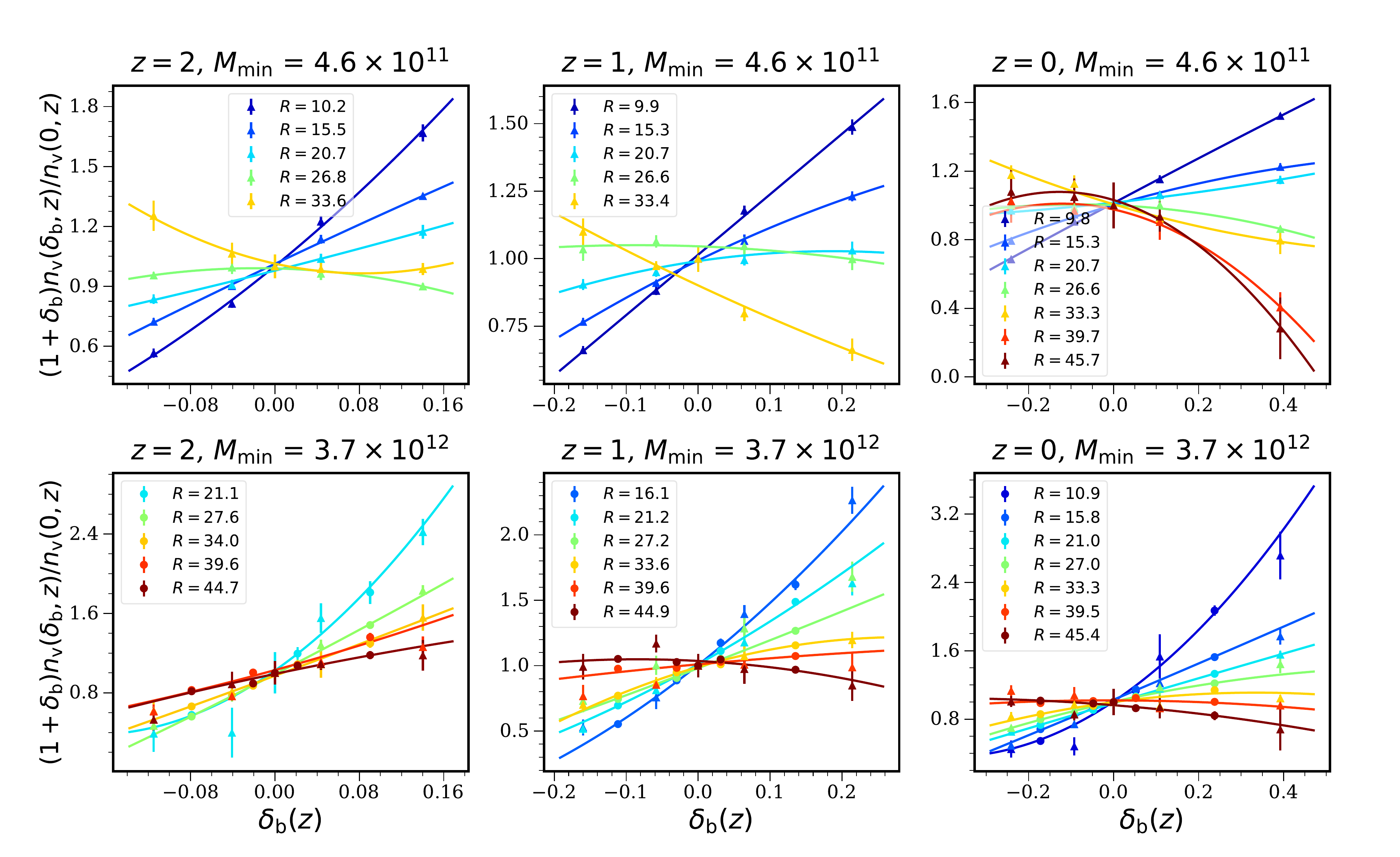}
\caption{ Similar to Fig.~\ref{fig:nv_DMvoids_deltab_Eul_subset}, but for halo voids obtained using the halo sample with minimum mass $4.6 \times 10^{11} $ (top panels, data from H-set only) and $3.7 \times 10^{12} \Msun$ (lower panels, data from both H-set and M-set).   }
\label{fig:nv_Halovoids_deltab_Eul_subset}
\end{figure*}

\begin{figure*}[!htb]
\centering
\includegraphics[width=0.9\linewidth]{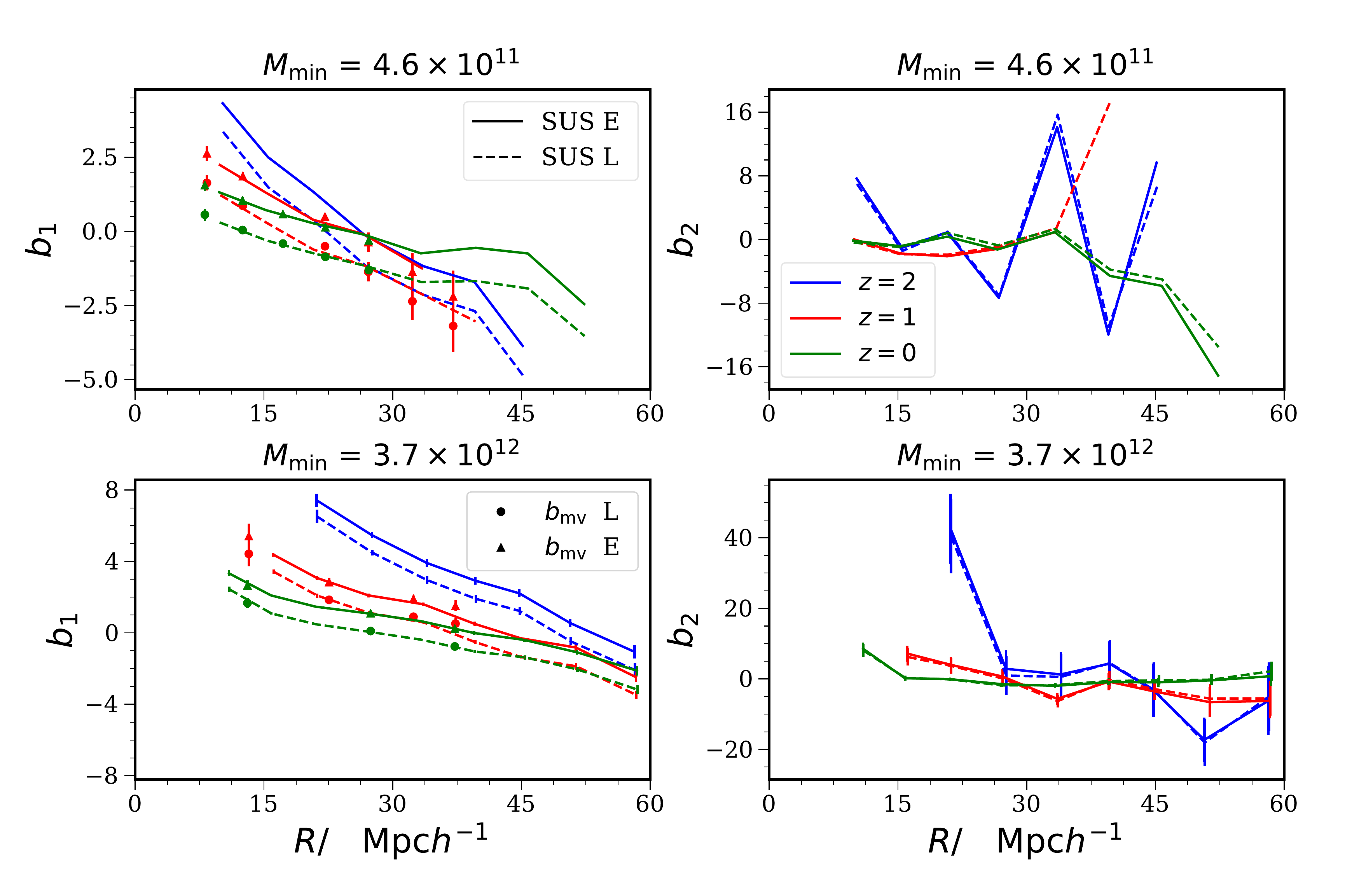}
\caption{  Similar to Fig.~\ref{fig:void_bias_DMvoid_subset}, but for halo voids obtained using the halo sample with minimum  mass $4.6 \times 10^{11} $ (top panels, data from H-set only) and $3.7 \times 10^{12} \Msun$ (lower panels, data from both H-set and M-set).  The signal to noise of the cross power spectrum measurements at $z=2$ is low and the error bars of the SUS results  for the $4.6 \times 10^{11} \Msun $ group cannot be reliably estimated (H-set data only), and so they are not shown.  }
\label{fig:void_bias_Halovoid_subset}
\end{figure*}


\begin{figure}[!htb]
\centering
\includegraphics[width=\linewidth]{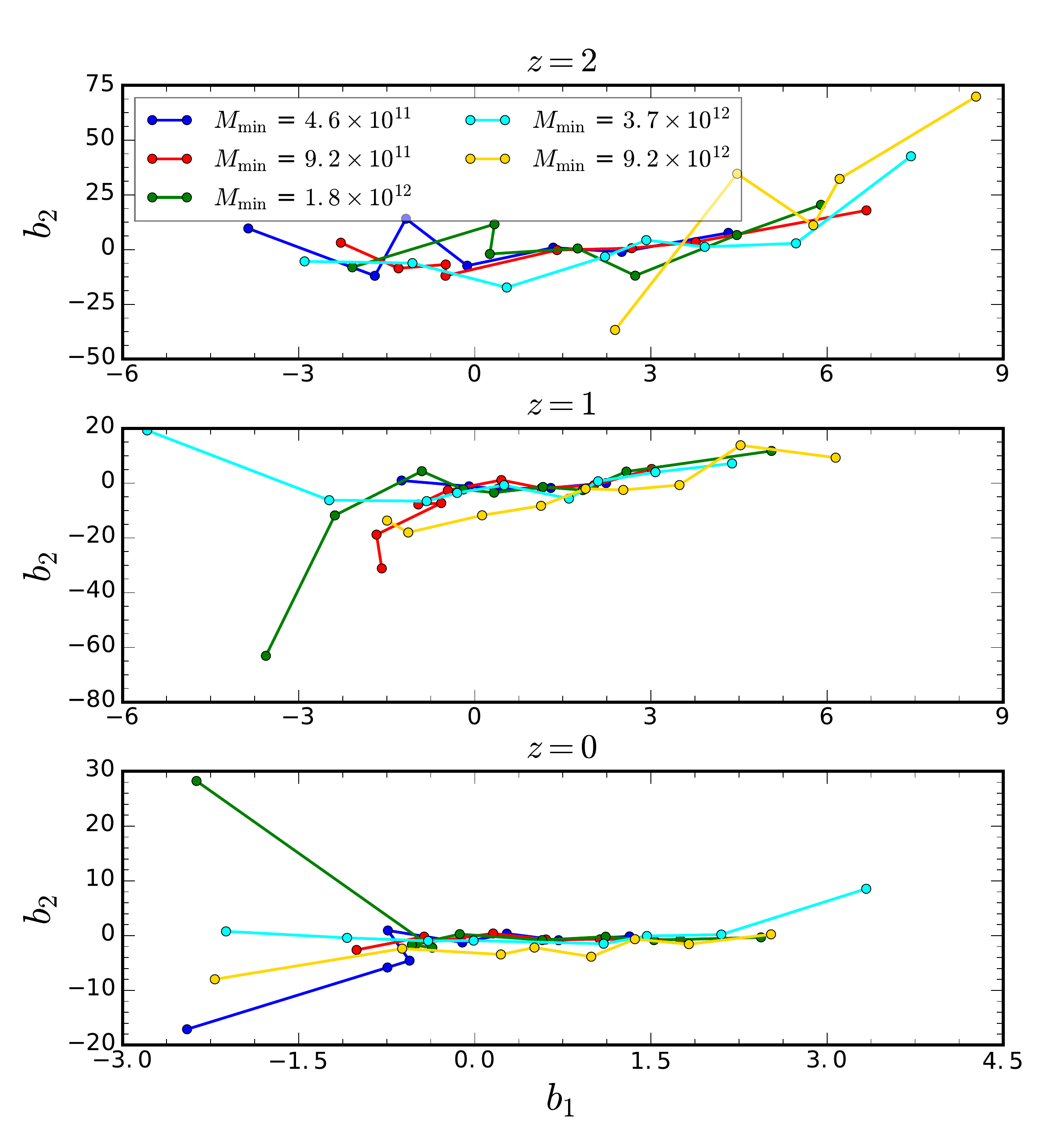}
\caption{ Best-fit SUS $b_2 $ versus $ b_1$ for  $z=2$, 1, and 0 (top to bottom panels).   The results are derived from the halo void sample with minimal mass of $4.6 \times 10^{11}$ (blue), $9.2 \times 10^{11}$ (red), $1.8 \times 10^{12} $ (green), $3.7 \times 10^{12} $ (cyan), and $9.2 \times 10^{12}  \Msun $ (yellow), respectively.  }
\label{fig:b1b2_Halovoids}
\end{figure}

\subsection{Halo void bias}

We now move to voids identified in the halo field. The voids from different SUS are constructed on the halo density field with the same minimal halo mass.  We show the void size distribution obtained with two different minimum halo mass thresholds ($4.6 \times 10^{11} $ and $  3.7 \times 10^{12} \Msun  $) in Fig.~\ref{fig:dnlnR_Halovoids_MHsets_subset}.  They correspond to the 20-particle threshold in the H-set and M-set respectively.  The response of the halo void size distribution is similar to the dark matter void case showing both the PR and NR regimes. 

In Fig.~\ref{fig:nv_Halovoids_deltab_Eul_subset}, we show the normalized halo void size distribution weighted by  $(1 + \delta_{\rm b})$  as a function of $\delta_{\rm b} $ for different void size bins. We fit a quadratic polynomial to extract the bias parameters, and the best-fit results are presented in Fig.~\ref{fig:void_bias_Halovoid_subset}. We also measure the large-scale bias from $b_{\rm mv} $ following the same procedure as for the dark matter void case. For $z=2$, the signal to noise of the cross bias measurement is low, and we do not show them here.  Similar to the dark matter void case, we find good agreement between $b_1$ from the SUS and the cross bias measurements. Again, we confirm consistency between the Eulerian and the Lagrangian results.

The detection of the quadratic bias is weak, especially for the low-threshold sample, which only has the H-set data. The high-threshold case suggests that the quadratic bias is positive in the PR region and  negative in the NR part.  For the PR part, positive $\delta_{\rm b} $ causes more breaking up into smaller regions.  The behaviour in the NR part is similar to the dark matter voids.


Fig.~\ref{fig:b1b2_Halovoids} shows the best-fit SUS $b_2$ versus $b_1$ for the halo voids. As $z$ decreases the spread of the data points in the $b_2-b_1$ plane reduces, and the data points for $z=0$ are concentrated in the central region of the plot.  As the minimum mass threshold increases, the data moves from the left part of the plot to the right part, but this trend becomes less and less apparent when the redshift decreases.  It is clear that the positive $b_1 $ part is associated with positive $b_2 $. The trend is not apparent for the negative  $b_1 $ part, but it is expected to be positive based on the dark matter void results.

\subsection{ Comparison between dark matter voids and halo voids }

In previous subsections, we looked at the bias parameters of the dark matter voids and the halo voids separately. In this subsection, we investigate the bias parameter of voids from matter and halo density fields with equal number density. Thus the difference between them is solely due to the tracer being biased or not. To do so, we choose the subsampling density for the dark matter voids  matching to the number density of the halo void sample.

\begin{figure}[!htb]
\centering
\includegraphics[width=\linewidth]{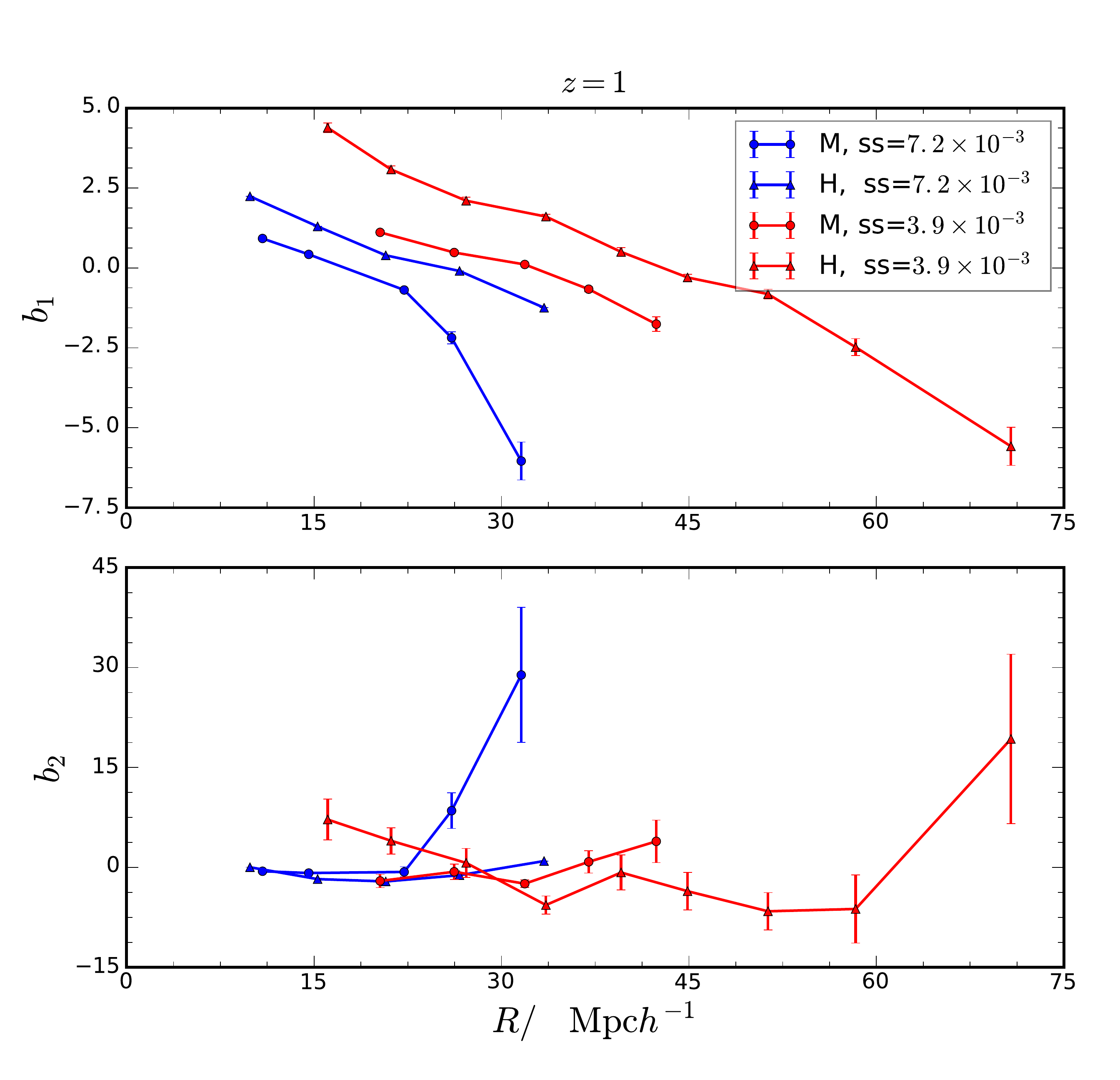}
\caption{  Eulerian bias parameters from the dark matter voids (circles) and the halo voids (triangles) with equal number density. Two subsampling densities are shown:  $3.9 \times 10^{-3 }$  and  $7.2 \times 10^{-3 }  (\MpcOh)^{-3}$.  The error bars of the halo void sample with  subsampling density $7.2 \times 10^{-3 }  (\MpcOh)^{-3}$ cannot be reliably estimated (H-set data only), and so they are not shown.
}
\label{fig:b1b2_DensityMatch}
\end{figure}

In Fig.~\ref{fig:b1b2_DensityMatch}, we compare the void bias obtained using samples at $ z=1 $ with the subsampling density of $3.9 \times 10^{-3 }$  and  $7.2 \times 10^{-3 }  (\MpcOh)^{-3}$.  For these halo voids, their minimum halo mass thresholds are $4.6 \times 10^{11} $ and $  3.7 \times 10^{12} \Msun  $ and their linear halo biases are 1.28 and 2.11, respectively. \revise{ The results for other redshifts are qualitatively similar, so we do not show them here. }

For the same subsampling density, $b_1$ from halo voids is higher than that from the dark matter voids. The halo void $b_2$  is lower than that from the dark matter voids in the NR regime, while it is higher than the dark matter void $b_2$ in the PR regime. \revise{ Alternatively, the shape of the halo void bias parameters is similar to that of the dark matter void ones, but shifted to larger void sizes.  This trend is  consistent with the finding in  \citet{Furlanetto:2005cc}   using the excursion set theory that halo (or galaxy) voids are generally larger in size than the dark matter voids. }

\subsection{ Further discussions }
\label{sec:further_discussions}

Overall, our results are consistent with the PBS argument, i.e.,~the void abundance is modulated by the long wavelength perturbations and the large-scale void bias can be derived by considering the effect of the long mode on the void abundance.  Our direct test using the SUS evades the need for a universal void size distribution and it tests the PBS at the fundamental level.  
\revise{  The void size distribution depends on the resolution via the number density of the tracers.  Furthermore, the voids identified in numerical algorithms such as the watershed algorithm do not necessarily correspond to the shell-crossing condition \citep{Blumenthal_etal1992} that is used to define voids theoretically.} The SUS bypass these difficulties by using the numerical void size distribution directly.

\revise{ There has been recent progress to clean the dark matter and halo-defined voids obtained from an arbitrary numerical definition so that the resultant catalogs are closer to the definition of voids in theory and hence yield a better agreement with theory predictions \citep{Jennings:2013nsa,Ronconi:2017gjc,Ronconi:2019xex, Verza:2019tvg, Contarini:2019qwf}.   However, these post processings substantially reduce the number density of voids, and this would increase the shot noise in the clustering measurement.  As our goal is to use voids as a tracer of the large-scale structure and, to a lesser extent, to adjust it to agree with a (well-motivated) theory, we used all the available voids in this work. }

\revise{ While our work is not tied to a specific void size distribution model, it is useful to have a theoretical model for void bias parameters. We compared the PBS bias model in \citet{Chan:2014qka} with the SUS bias measurements, treating the void formation threshold $\dv $ as a fitting parameter (see also \citet{PisaniSutter_etal2015}). The model is based on a simple first crossing distribution for voids given in \citet{Sheth:2003py}  and it is aimed for the dark matter voids. It was found that to obtain a qualitative agreement with the numerical results the Eulerian void radius has to be scaled down by a factor of a few. This is not surprising, because Fig.~2 of \citet{Ronconi:2017gjc} shows that many of the {\tt VIDE} voids actually correspond to theoretical voids of smaller sizes.  Thus, we  need to apply the cleaning procedures to the {\tt VIDE} voids first before comparing with the model.  Besides cleaning, it is also desirable to include the resolution dependence in the modeling. }  

The linear void bias is often obtained from the correlation measurement, especially the cross power spectrum on large scale. However, it is not always clear that $ b_{\rm mv} $ has reached a constant. This can be because the data is noisy or the scale of the measurement is not large enough so that the higher-order bias parameters and/or void profile are negligible.  In the previous section, we use the cross power spectrum to verify the SUS results. However, it is also useful to turn the argument around and assume that the large-scale linear bias is given by the SUS result and check how large the scale is required to get a reliable measurement of $b_1$ from  the cross power spectrum.

\begin{figure*}[!htb]
\centering
\includegraphics[width=0.9\linewidth]{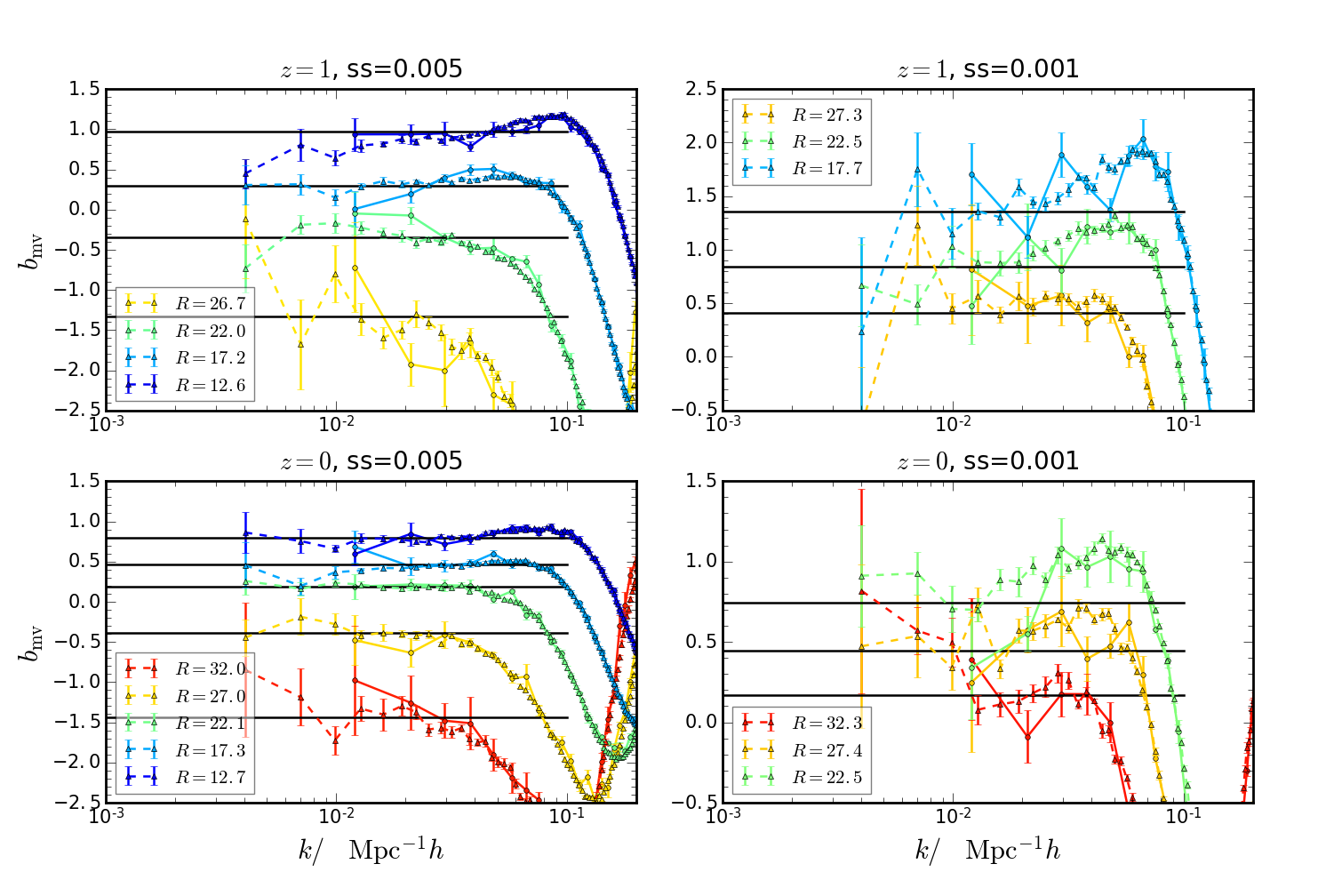}
\caption{  Cross bias $b_{\rm mv}$ (triangles from L-set and circles from M-set) and the $b_1 $ from the SUS (black lines). The dark matter void samples at $z=1$ and 0 (top and bottom panels) obtained with subsampling density 0.005 and 0.001 $(\MpcOh)^{-3}$ (left and right panels) are used.  }
\label{fig:bmv_Pc_SUS_subset}
\end{figure*}

In Fig.~\ref{fig:bmv_Pc_SUS_subset}, we compare the cross bias $b_{\rm mv}  $ against the SUS  $b_1$.  We show the results from the dark matter voids at $z=1$ and 0.  In order to see the large-scale bias more clearly, we have plotted the $b_{\rm mv} $ measurement from the L-set in addition to  the M-set ones. Overall, the SUS results are in good agreement with $ b_{\rm mv} $ within the uncertainties for $k$ smaller than $0.04 \hOMpc$. \revise{ The error bars are derived from the standard deviations of the measurements  among the realizations (six and eight, respectively).}  However, we note that for the smallest voids shown, $ b_{\rm mv} $ appears to be moderately scale dependent even on large scales. The void size bin $ R=12.6 \MpcOh$ at $z=1$ with the subsampling density $0.005 (\MpcOh)^{-3} $ is a typical example. It is not likely that the discrepancy is due to the quadratic bias, as its magnitude is similar to the other bin sizes.   This void size bin corresponds to the PR zone, which becomes steeper as the subsampling density increases and as the redshift increases (c.f.~Fig.~\ref{fig:dnlnR_DMvoids_MHsets_subset}). It is numerically challenging to capture the small change in the void abundance in this case. Thus, the discordance can be attributed to the insufficient resolution in the SUS.  Note that in this particular case, the SUS result agrees with the $b_{\rm mv}$ from M-set well. These indicate that the voids in the PR regime (especially at high redshift) requires particularly large simulation volume to simulate them well.  In the comparison between the SUS and $ b_{\rm mv} $ results in Fig.~\ref{fig:void_bias_DMvoid_subset} and \ref{fig:void_bias_Halovoid_subset}, we have avoided the pathological cases in the PR regime.

\section{Conclusions}
\label{sec:conclusion}

The SUS furnish an accurate implementation of the PBS. We have applied the SUS technique to test the PBS argument for  underdense tracers at the fundamental level and measure the resultant void bias parameters. 

We have considered both the dark matter voids and the halo voids, and their response to a long mode is qualitatively similar. We identify two regimes of the void size distribution response, the NR and the PR regimes. In the NR regime, the void abundance decreases with $ \delta_{\rm b} $.  These voids correspond to large underdense regions, and an increase of $ \delta_{\rm b} $ makes it harder to form voids. These large underdense regions are relatively stable with respect to  $ \delta_{\rm b} $.  In the PR regime, the void abundance instead increases with $ \delta_{\rm b} $. A physical scenario is that when $\delta_{\rm b} $ increases, an initially smooth region breaks up into a few collapsed structures and this segmentation process creates shallow zones in between.  Hence, in this scenario, voids are by-products of the collapse of the other structures. These two scenarios are mixed in reality, but they play a more important role in one regime than the other.

From the SUS we measure the linear void bias and for the first time the quadratic void bias. We have checked that the SUS linear bias measurements are in \revise{good} agreement with the results from the cross power spectrum between matter and voids.  The results confirm the consistency between the Lagrangian and the Eulerian bias for voids. In particular, the relation between the linear Lagrangian bias and the Eulerian one implies that voids co-evolve with the underlying dark matter density field.  The halo void biases are generally larger than the dark matter void ones and extend to larger void sizes. When the magnitude of $b_1$ is relatively large, $b_2$ becomes positive.

A better theoretical void size distribution is needed. Among the two regimes,  the NR part of the distribution is relatively easy to model, as it fits well with the standard excursion set theory picture, but the PR regime will be more difficult to model because it requires the interplay between voids and halos. After all, even the low-mass part of the halo mass function proves difficult for the excursion theory to model accurately. \revise{ To properly compare it  with theoretical models, cleaning strategies, such as that advocated by \citet{Jennings:2013nsa,Ronconi:2017gjc} can be used.}  On the other hand, as the SUS enable us to have a handle on the quadratic bias, it will facilitate other studies of voids such as the assembly bias of voids  and their Lagrangian profile evolution.

\revise{Shortly after this draft was submitted, \citet{JamiesonLoverde2019} appeared on the arXiv and also reported the measurement of the void bias using the SUS.  Their linear bias measurements are consistent with ours. Unlike us, they used the linear SUS and do not investigate the quadratic bias. Instead, they present a decomposition of different terms that enter the linear bias, and investigate void profiles within the SUS framework.}

\section*{Acknowledgements}

We thank Vincent Desjacques, Aseem Paranjape, and especially Ravi Sheth for useful discussions. \revise{We also thank the anonymous referee for the constructive comments that improved the presentation of the manuscript.  } The SUS were run at the Kunlun cluster at the School of Physics and Astronomy of the Sun-Yat Sen University. We thank the system administrator Yang Wang and Weishan Zhu for help with the cluster.  KCC acknowledges the support from the National Science Foundation of China under the grant 11873102.  YL acknowledges support from Fellowships at the Kavli IPMU established by World Premier International Research Center Initiative (WPI) of the MEXT, Japan, and at the Berkeley Center for Cosmological Physics. MB acknowledges support from the Netherlands Organisation for Scientific Research (NWO), which is funded by the Dutch Ministry of Education, Culture and Science (OCW), under VENI grant 016.Veni.192.210.”  This research was supported by the Munich Institute for Astro- and Particle Physics (MIAPP) of the DFG cluster of excellence ``Origin and Structure of the Universe.''

\software{2LPTic ~\citep{CroccePeublasetal2006},  Gadget2~\citep{Gadget2},  AHF~\citep{AHF_2009},  ZOBOV~\citep{Neyrinck2008},  VIDE~\citep{Sutter:2014haa}, python, numpy, scipy }

\appendix

\section{ Lagrangian to Eulerian space mapping   }
\label{sec:LagEul_mapping}

Let us first review the derivation of the Eulerian bias in Eq.~\eqref{eq:bi_Eulerian}. Suppose there are $N$ objects (halos or voids) in a comoving volume, $V_{\rm L}$ and its number density is $ n_{\rm L}$ in Lagrangian space. If the number of objects conserves during the collapse/expansion phase, the volume occupied by the objects in Eulerian space is
$V_{\rm E}$, then the number density in Eulerian space $ n_{\rm E} $ reads
\beq
\label{eq:nE_Nconserve}
 n_{\rm E} = \frac{ V_{\rm L}  }{ V_{\rm E} }  n_{\rm L}  =  ( 1 + \delta_{\rm E} ) n_{\rm L}   .
 \eeq
 The factor  $ 1 + \delta_{\rm E} $ arises from conservation of mass on large scale.
The spatial dependence in Eulerian space can be expressed in terms of the Eulerian bias parameters as
\begin{align}
  n_{\rm E} &= \bar{n} \left(1 + b_1 \delta_{\rm E } + \frac{1 }{2 } b_2 \delta_{\rm E }^2  + \dots     \right) ,
\end{align}
and similarly in Lagrangian space
\beq
n_{\rm L} = \bar{n}  \left( 1 +  b_1^{\rm L } \delta_{\rm L} + \frac{1}{2} b_2^{\rm L } \delta_{\rm L}^2 + \dots \right).
\eeq
Note that $\delta_{\rm E} $ is $ \delta_{\rm b} $ and  $\delta_{\rm L} $ is  $ \delta_{\rm b}^{\rm L} $ in the main text. Hence the Eulerian bias parameter $b_i$ is given by [i.e.~Eq.~\eqref{eq:bi_Eulerian}]
\beq
\label{eq:bi_Eulerian_1}
b_i = \frac{ \partial^i }{ \partial \delta^i_{\rm E} } \left[  (1 + \delta_{\rm E} ) \frac{ n_{\rm L}( \delta_{\rm L} ) }{ \bar{n}  } \right] \bigg|_{\delta_{\rm E } =0  }.
\eeq
We can use spherical collapse to relate $\delta_{\rm L}$ to $\delta_{\rm E} $ \citep{Bernardeau1994}
\beq
\delta_{\rm E} = \delta_{\rm L} + \alpha_1 \delta_L^2  + \alpha_2 \delta_{\rm L}^3  \dots,
\eeq
where $\alpha_1 = 17/21$ and  $\alpha_2 =341/567 $.  The number density in Eq.~\eqref{eq:bi_Eulerian_1} can be replaced by the mass function or the void size distribution. The Eulerian bias parameters are related to the Lagrangian ones  as
\begin{align}
  \label{eq:b1_MassConserve}
  b_1 &= 1 + b_{1}^{\rm L} \\
    \label{eq:b2_MassConserve}
  b_2 &= 2 ( 1- \alpha_1)  b_1^{\rm L} + b_2^{\rm L} .
\end{align}

On the other hand, \citet{Jennings:2013nsa} noted that if the number density of voids conserves, after the spherical expansion, the volume occupied by voids predicted by the original void size distribution of \citet{Sheth:2003py}  would exceed the volume of the universe. To cure this inconsistency,  they proposed that the total volume occupied by the voids does not change instead they merge to become voids of different sizes, i.e.,
\beq
\bar{n}_{\rm L} \mathcal{V}_{\rm L} = \bar{n}_{\rm E} \mathcal{V}_{\rm E} = \mathrm{const},
\eeq
where  $\mathcal{V}_{\rm L}$  and  $\mathcal{V}_{\rm E} $ are the volumes occupied by voids in the Lagrangian and the Eulerian space, respectively. To be consistent with our bias measurement, in this model the total volume occupied by voids in $V_{\rm L}$ and $V_{\rm E} $ are equal to each other, i.e.,
\beq
n_{\rm E} = \frac{ V_{\rm L} }{ V_{\rm E}} \frac{ \mathcal{V}_{\rm L} }{  \mathcal{V}_{\rm E}  } n_{\rm L}.
\eeq
In the simple spherical collapse model, the factor $ \mathcal{V}_{\rm L} / \mathcal{V}_{\rm E} \approx 1 / 5 $. Note that even if the voids conserve volume among themselves, we still need the total mass to be conserved in the mapping from $V_{\rm L}$ to $V_{\rm E} $.  Expanding $ n_{\rm E}$ and  $n_{\rm L}$ about their mean values, we recover also Eqs.~\eqref{eq:b1_MassConserve} and \eqref{eq:b2_MassConserve}.

Eq.~\eqref{eq:b1_MassConserve} holds whenever voids move along with the large-scale structure regardless the details of the  void formation physics.  Our measurements clearly support Eq.~\eqref{eq:b1_MassConserve} [see also \citet{Massara:2018dqb}].  These two different perspectives can be classified as theory-oriented and algorithm-oriented. In the theory-oriented approach, one takes a theory for voids in Lagrangian space, and changes the evolution to agree with the Eulerian results.  The volume-conserving model falls into this category. The model is incomplete until further prescriptions to specify how voids merge and keep the volume constant in the Lagrangian space to Eulerian space mapping.  In the algorithm-oriented approach, the Lagrangian halos/voids are constructed by tracing the particles in the Eulerian space back to the Lagrangian one regardless whether merging has undergone or not.  The numerical studies in this case are much simplified as there is no need for the merger tree, and these objects conserve mass by construction. This is often adopted in studying the Lagrangian halos, e.g.~\citet{ChanShethScoccimarro2017_1,ChanShethScoccimarro2017_2}.  The job is then boiled down to finding a proper model that can explain the Lagrangian results. We have taken the algorithm-oriented perspective in the main text.

\section{Abundance matching to second order}
\label{sec:ABmatching}

In \citet{LiHuTakada2016}, rather than directly measuring the halo bias from the halo mass  function as a function of $\delta_{\rm b} $ as the method presented in the main text, they developed the so-called abundance matching method. In this method, one first ranks the halos starting from the largest mass, and then matches their rank assuming there is a one-to-one correspondence between the halos of the same rank from the SUS with different $\delta_{\rm b}$.  \citet{LiHuTakada2016} used this method for the linear bias measurement,  here we extend the abundance matching technique to measure the quadratic bias.

We first define the cumulative abundance function as
\begin{align}
\mathcal{N} (M_{\rm th} , \delta_{\rm b}) = \int_{ M_{\rm th}}^\infty  d \ln M  \, \mathfrak{n}(M, \delta_{\rm b} ) ,
\end{align}
where $\mathfrak{n}  $ may denote the halo mass function or the void size distribution.  The threshold  $M_{\rm th}$ ($R_{\rm th}$ for the case of void size distribution)  is defined such that the cumulative abundance $\mathcal{N}$ is independent of $\delta_{\rm b} $. Consequently, we have
\beq
\label{eq:Mth_1derivative_condition}
\frac{ d^i \mathcal{N} }{ d \delta_{\rm b}^i } = 0,
\eeq
for any $i$.

The first and second derivatives of $\mathcal{N} $ read
\begin{align}
  \frac{ d \mathcal{N} }{ d \delta_{\rm b} } & = \int_{M_{\rm th}}^\infty d \ln M \frac{ \partial \mathfrak{n}  }{\partial \delta_{\rm b} } - \frac{d \ln M_{\rm th} }{d \delta_{\rm b} } \mathfrak{n}(M_{\rm th}, \delta_{\rm b} ) , \\
  \frac{ d^2 \mathcal{N} }{ d \delta_{\rm b}^2 } & = \int_{M_{\rm th}}^\infty d \ln M \frac{ \partial^2 \mathfrak{n} (M, \delta_{\rm b}) }{\partial \delta_{\rm b}^2 }   -    2 \frac{d \ln M_{\rm th} }{d \delta_{\rm b} } \frac{ \partial \mathfrak{n} (M_{\rm th}, \delta_{\rm b}) }{\partial \delta_{\rm b} }
    -   \frac{d^2 \ln M_{\rm th}  }{d \delta_{\rm b}^2 } \mathfrak{n}  (M_{\rm th}, \delta_{\rm b})
	- \biggl(\frac{d \ln M_{\rm th} }{d \delta_{\rm b} }\biggr)^2 \frac{ \partial \mathfrak{n} (M_{\rm th}, \delta_{\rm b})  }{\partial \ln M_{\rm th} }.
 \end{align}
Evaluating Eq.~\eqref{eq:Mth_1derivative_condition} for $i=1$ and 2 at $ \delta_{\rm b} =0 $, we get
\begin{align}
    \int_{M_{\rm th}}^\infty d \ln M \,  b_1   \mathfrak{n}  & = \frac{d \ln M_{\rm th} }{d \delta_{\rm b} } \mathfrak{n}(M_{\rm th} , 0) ,  \\
  \int_{M_{\rm th}}^\infty d \ln M  b_2 \mathfrak{n} &  =   2 b_1  \mathfrak{n}  \frac{d \ln M_{\rm th} }{d \delta_{\rm b} }   +
  \frac{d^2 \ln M_{\rm th} }{d \delta_{\rm b}^2 } \mathfrak{n}
   +   \biggl(\frac{d \ln M_{\rm th} }{d \delta_{\rm b} }\biggr)^2 \frac{ \partial \mathfrak{n} (M_{\rm th},0)  }{\partial \ln M_{\rm th} } .
\end{align}
We now define the bin-averaged linear and quadratic biases $[b_1]$ and $[b_2]$ as
\begin{align}
  [b_1]  &\equiv \frac{ \int_{M_1}^{M_2} d \ln M \,  b_1  \mathfrak{n}  }{ \int_{ M_1}^{ M_2} d \ln M \, \mathfrak{n}   }
  =  \frac{ \left[  \mathfrak{n}  \frac{d \ln M_{\rm th} }{d \delta_{\rm b} } \right]_{M_2}^{M_1}  }{  \int_{ M_1}^{ M_2} d \ln M \, \mathfrak{n}   },  \\
    [b_2]  &\equiv  \frac{ \int_{M_1}^{M_2} d \ln M \,  b_2  \mathfrak{n}  }{ \int_{ M_1}^{ M_2} d \ln M \, \mathfrak{n}   }
          =  \frac{ \left[ 2 b_1 \mathfrak{n} \frac{d \ln M_{\rm th} }{d \delta_{\rm b} } + \mathfrak{n}  \frac{d^2 \ln M_{\rm th} }{d \delta_{\rm b}^2 }  +   \biggl(\frac{d \ln M_{\rm th} }{d \delta_{\rm b} }\biggr)^2 \frac{ \partial n (M_{\rm th}, \delta_{\rm b})  }{\partial \ln M_{\rm th} }        \right]_{M_2}^{M_1}  }{     \int_{M_1}^{M_2} d \ln M \, \mathfrak{n}  },
\end{align}
where $M_1$ and $M_2$  are the boundaries of the mass bin.

To get  $ \frac{d \ln M_{\rm th} }{d \delta_{\rm b} } $ and  $\frac{d^2 \ln M_{\rm th} }{d \delta_{\rm b}^2 }  $, we  match $ M_{\rm th} $ of the same rank and then fit a quadratic polynomial in $\delta_{\rm b} $ to it. The coefficients of the polynomial would give these functions, and so they are similar to the direct method in spirit. In addition, to the second order, the logarithmic derivative of the mass function is also required. However, we find that in the abundance matching method random noise can give rise to some non-local correlation. Suppose there is some fluctuations at $M$ in one of the SUS, by matching the rank, the fluctuation would cause all the ranks subsequent to $M$ to be displaced.   Consequently, the noise causes long range correlation. Indeed we find that the quadratic bias measurement suffers from some spurious oscillations. Because these oscillations are smooth, it is hard to disentangle them from the real signal. It is relatively mild for the halo case, but for voids the signal-to-noise is low, the spurious oscillations are strong. Thus, in the main text, we opt to use the direct method.

\clearpage
\bibliographystyle{aasjournal}
\bibliography{reference}



\end{document}